# Universal Linear Intensity Transformations Using Spatially-Incoherent Diffractive Processors

Md Sadman Sakib Rahman[1,2,3]   mssr@ucla.edu
Xilin Yang[1,2,3]                mikexlyang@ucla.edu
Jingxi Li[1,2,3]                 jxlli@ucla.edu
Bijie Bai[1,2,3]                 baibijie@ucla.edu
Aydogan Ozcan[1,2,3,*]            ozcan@ucla.edu

[1]Electrical and Computer Engineering Department, University of California, Los Angeles, CA, 90095, USA
[2]Bioengineering Department, University of California, Los Angeles, CA, 90095, USA
[3]California NanoSystems Institute (CNSI), University of California, Los Angeles, CA, 90095, USA
[*]Corresponding author: ozcan@ucla.edu

## Abstract

Under spatially-coherent light, a diffractive optical network composed of structured surfaces can be designed to perform any arbitrary complex-valued linear transformation between its input and output fields-of-view (FOVs) if the total number ($N$) of optimizable phase-only diffractive features is $\geq \sim 2N_i N_o$, where $N_i$ and $N_o$ refer to the number of useful pixels at the input and the output FOVs, respectively. Here we report the design of a spatially-incoherent diffractive optical processor that can approximate any arbitrary linear transformation in time-averaged intensity between its input and output FOVs. Under spatially-incoherent monochromatic light, the spatially-varying intensity point spread function ($H$) of a diffractive network, corresponding to a given, arbitrarily-selected linear intensity transformation, can be written as $H(m,n;m',n') = |h(m,n;m',n')|^2$, where $h$ is the spatially-coherent point-spread function of the same diffractive network, and $(m,n)$ and $(m',n')$ define the coordinates of the output and input FOVs, respectively. Using deep learning, supervised through examples of input-output profiles, we numerically demonstrate that a spatially-incoherent diffractive network can be trained to all-optically perform any arbitrary linear intensity transformation between its input and output if $N \geq \sim 2N_i N_o$. These results constitute the first demonstration of universal linear intensity transformations performed on an input FOV under spatially-incoherent illumination and will be useful for designing all-optical visual processors that can work with incoherent, natural light.

## Introduction

Spatial information processing with free-space optics has been widely explored for a long time and predates the proliferation of electronic computing [1–18]. The inherent transformation of optical fields as they propagate through free space, known as diffraction, together with the ability for wavefront modulation with compact hardware, makes low-cost and passive spatial information processing at the speed of light propagation possible [19–21]. In recent years, diffractive optical networks comprising a set of spatially-engineered surfaces to perform computation through passive light-matter-interaction have emerged as powerful all-optical processors [22,23]. Designed utilizing deep learning [24], such coherent diffractive optical processors have demonstrated versatile applications, including statistical inference as well as deterministic tasks[23,25–31] across the spectrum from terahertz to near-infrared[32] and visible[33,34].

Information processing with a diffractive network involves local modulation of the amplitude and/or the phase of the incident optical wave by structured surfaces containing diffractive neurons/features, each with a lateral size of ~$\lambda/2$, where $\lambda$ is the wavelength of the spatially-coherent illumination light. The entire propagation of a spatially-coherent wave from the input plane to the output FOV comprises such optical modulations by $K$ spatially-optimized diffractive surfaces, which in total contain $N$ independent diffractive features (for example, evenly distributed over the $K$ diffractive surfaces). These $N$ diffractive features represent the complex-valued transmission coefficients, forming the independent degrees of freedom of the diffractive processor, which can be optimized to all-optically execute different tasks[22,26–30,35,36]. It was shown that a spatially-coherent diffractive optical network could be trained to perform any arbitrary complex-valued linear transformation between its input and output FOVs if $N \geq N_i N_o$, where $N_i$ and $N_o$ refer to the number of useful (diffraction-limited) pixels at the input and the output FOVs [21]. For a phase-only diffractive network where the transmission coefficients of the diffractive features of each structured surface only modulate the phase information of light, the requirement for universal linear transformations increases to $N \geq 2 N_i N_o$ due to the reduced degrees of freedom that can be optimized independently.

For a given complex-valued linear transformation that a coherent diffractive network is designed to approximate, any arbitrary point on the input plane defined by $(m', n')$ will result in a unique complex-valued coherent point spread function ($h$) at the output FOV defined by $(m, n)$. This 4-dimensional complex-valued function, $h(m, n; m', n')$, that maps the input and output FOVs represents a *spatially-varying* coherent point spread function. Stated differently, unlike traditional spatially-invariant imaging systems, a coherent diffractive optical network provides a framework to approximate any arbitrary $h(m, n; m', n')$ that corresponds to an arbitrarily-selected complex-valued linear transformation between its input and output FOVs. It was also shown that different/independent complex-valued linear transformations could be multiplexed in a single spatially-coherent diffractive processor by utilizing polarization and wavelength diversity[37,38].

All of these earlier studies on universal linear transformations implemented in free-space through diffractive processors were based on spatially-coherent illumination. In this paper, we report the first demonstration of universal linear transformations in optical intensity performed under spatially-incoherent monochromatic illumination of an input FOV. We show that, under spatially-incoherent light, a diffractive optical processor can perform any arbitrary linear transformation of time-averaged intensities between its input and the output FOVs. Our numerical analyses revealed that phase-only diffractive optical processors with a shallow architecture (for example, having a single trainable diffractive surface) are unable to accurately approximate an arbitrary intensity transformation irrespective of the total number ($N$) of diffractive features available for optimization; on the contrary, phase-only diffractive optical processors with deeper architectures (one diffractive layer following others) can perform an arbitrary intensity linear transformation using spatially-incoherent illumination with a negligible error when $N \geq 2 N_i N_o$. These analyses and conclusions are important for all-optical information processing and visual computing systems that use spatially-incoherent light, such as in natural scenes; this framework can also find applications in computational microscopy and incoherent imaging through point spread function engineering, among others.

## Results

### Theoretical analysis

Spatially-coherent monochromatic diffractive optical networks can be characterized by a 4-dimensional complex-valued coherent impulse response function (i.e., the point spread function) that is spatially-varying, connecting the input and output FOVs: $h(x, y; x', y')$. Stated differently, each arbitrarily-selected complex-valued linear transformation that is desired between the pixels of an input FOV and output FOV results in a spatially-varying impulse response function $h(x, y; x', y')$, where $(x', y')$ and $(x, y)$ define the input and output FOVs, respectively. Based on this definition, the complex-valued output field $o_c(x, y)$ of a spatially-coherent diffractive processor is related to the complex-valued input field $i_c(x', y')$ by:

$$o_c(x, y) = \iint h_c(x, y; x', y') i_c(x', y') dx' dy' \quad (1)$$

The subscript $c$ indicates that the quantities are functions of continuous spatial variables $x, y, x', y'$, representing the transverse coordinates on the output and input planes. If these optical fields are sampled at an interval ($\delta$) sufficiently small to preserve the spatial variations, satisfying the Nyquist criterion[39], one can write:

$$o(m, n) = \sum_{m',n'} h(m, n; m', n') i(m', n') \quad (2)$$

Here, $m, n, m', n'$ refer to discrete indices such that $o(m, n) = o_c(m\delta, n\delta)$ and $i(m', n') = i_c(m'\delta, n'\delta)$. The instantaneous output intensity can be written as:

$$|o(m,n)|^2 = \sum_{m',n',m'',n''} h(m,n;m',n') h^*(m,n;m'',n'') |i(m',n')| |i(m'',n'')| e^{j(\varphi(m',n') - \varphi(m'',n''))} \quad (3)$$

where $\varphi(.)$ is the phase function of the input field $i$, i.e., $i = |i|e^{j\varphi}$, and $h^*$ denotes the complex conjugate of $h$. The time-averaged output intensity can be written as:

$$O(m,n) = \langle |o(m,n)|^2 \rangle = \sum_{m',n',m'',n''} h(m,n;m',n') h^*(m,n;m'',n'') |i(m',n')||i(m'',n'')|\langle e^{j\Delta\varphi}\rangle \quad (4)$$

where $\langle \cdot \rangle$ denotes time-average operation and $\Delta\varphi = \varphi(m', n') - \varphi(m'', n'')$. Since the illumination light is spatially-incoherent, the phases at different spatial points of the input vary randomly over time and are independent of each other.[40] Stated differently for stationary objects/scenes that are uniformly illuminated with a spatially-incoherent light, $\Delta\varphi$ varies randomly between 0 and $2\pi$ over time, yielding $\langle e^{j\Delta\varphi}\rangle = 0$ for $(m', n') \neq (m'', n'')$. As a result of this, under spatially-incoherent illumination, Eq. (4) can be written as:

$$O(m,n) = \sum_{m',n'} |h(m,n;m',n')|^2 \langle |i(m',n')|^2 \rangle = \sum_{m',n'} H(m,n;m',n') I(m',n') \quad (5)$$

where $I = \langle |i|^2 \rangle$ is the time-averaged input intensity and $H(m, n; m', n') = |h(m, n; m', n')|^2$ is the intensity impulse response of the diffractive optical processor under spatially-incoherent illumination. From now on, unless otherwise stated, we use the term optical 'intensity' to imply time-averaged intensity functions. Similarly, whenever all-optical linear transformation of intensity is mentioned, spatially-incoherent monochromatic illumination is implied unless stated otherwise.

We should emphasize that while $H(m, n; m', n') = |h(m, n; m', n')|^2$, we have in general $O(m, n) \neq |o(m, n)|^2$. Therefore, the output intensity of a spatially-incoherent diffractive network cannot be calculated as $|o(m, n)|^2 = \left|\sum_{m',n'} h(m, n; m', n') \, i(m', n')\right|^2$. For the numerical forward model corresponding to each input object, as will be detailed in the next section, we used a large number of random phase distributions at the input plane to approximate $O(m, n) = \langle |o(m, n)|^2 \rangle$ under spatially-incoherent illumination.

## Numerical analysis

In this subsection, we numerically explore the design of diffractive optical processors to perform an arbitrary linear intensity transformation between the input and the output FOVs under spatially-incoherent illumination. We assume, as shown in Fig. 1a, $N$ independent diffractive features (phase-only elements) that are distributed over $K$ diffractive surfaces, each with $N/K$ diffractive features, between the input and output planes. Following from Eq. (5), if we rearrange the pixel intensities of $I(m', n')$ and $O(m, n)$ as column vectors $\boldsymbol{i}$ and $\boldsymbol{o}$, then we can write $\boldsymbol{o} = \boldsymbol{A}' \, \boldsymbol{i}$, where $\boldsymbol{A}'$ represents the linear intensity transformation performed by the diffractive optical network under spatially-incoherent illumination. The elements of $\boldsymbol{A}'$ correspond to the elements of the intensity impulse response $H(m, n; m', n')$; see Eq. (5). Note that all the elements of $\boldsymbol{A}'$ are real and nonnegative since it represents a linear intensity transformation with $H(m, n; m', n') = |h(m, n; m', n')|^2$. Hence, in the context of arbitrary linear transformations in intensity, only real transformation matrices with nonnegative elements are considered.

For our target linear transformation that is to be approximated by the spatially-incoherent diffractive processor, initially, we selected an arbitrary matrix $\boldsymbol{A}$, as shown in Fig. 1b. In the following numerical analysis, we optimize $N$ diffractive features of a phase-only diffractive processor so that $\boldsymbol{A}' \approx \boldsymbol{A}$ under spatially-incoherent illumination. The size of $\boldsymbol{A}$ is chosen as $N_o \times N_i = 64 \times 64$, i.e., the number of pixels at both the input ($N_i$) and the output ($N_o$) FOVs are $8 \times 8$, arranged in a square grid. Each element of the matrix $\boldsymbol{A}$ is randomly sampled from a uniform probability distribution between 0 and 1, i.e., $\boldsymbol{A}[p, q] \sim Uniform(0, 1)$ where $\boldsymbol{A}[p, q]$ is the element at $p$-th row and $q$-th column of $\boldsymbol{A}$, $p = 1, \ldots, N_o$ and $q = 1, \ldots, N_i$.

For the deep learning-based optimization of the design of a phase-only diffractive processor to achieve $\boldsymbol{A}' \approx \boldsymbol{A}$, we followed two different data-driven supervised learning approaches: (1) *the indirect approach* and (2) *the direct approach*. In the indirect approach, instead of directly training the diffractive network to perform the linear intensity transformation $\boldsymbol{A}$, we trained the network, under *spatially-coherent* illumination, to perform the complex-valued linear transformation $\bar{\bar{\boldsymbol{A}}}$ between the input and output FOVs such that $|\bar{\bar{\boldsymbol{A}}}[p, q]| = \sqrt{\boldsymbol{A}[p, q]}$, which would result in an intensity linear transformation $|\bar{\bar{\boldsymbol{A}}}[p, q]|^2 = \boldsymbol{A}[p, q]$ under spatially-incoherent illumination. For the purpose of the training, we defined the phase of $\bar{\bar{\boldsymbol{A}}}[p, q]$ to be zero, i.e., $\bar{\bar{\boldsymbol{A}}}[p, q] = \sqrt{\boldsymbol{A}[p, q]} \exp(j0)$; however, any other phase distribution could also be used since the design space is not unique. Stated differently, in this indirect approach, we design a diffractive network that can achieve a spatially-coherent impulse response $h(m,n;m',n')$, which will ensure that the same design has a spatially-incoherent impulse response of $H(m,n;m',n') = |h(m,n;m',n')|^2$ such that $A' \approx A$ can be satisfied under spatially-incoherent illumination. To achieve this goal, we used the relationship $\tilde{o} = \bar{\bar{A}}\tilde{i}$ to generate a large set of input-target complex-valued optical field pairs $(\tilde{i}, \tilde{o})$, and used deep learning to optimize the phase values of the diffractive features by minimizing the mean squared error (MSE) loss between the target complex field $\tilde{o}$ and the complex field $\tilde{o}'$ obtained by coherently propagating $\tilde{i}$ through the diffractive network (see the Methods section). In other words, spatially-coherent design of a diffractive network is used here as a proxy for the design of a spatially-incoherent diffractive network that can achieve any arbitrary intensity linear transformation between its input and output FOVs.

In the second approach (termed the direct approach), we trained the diffractive network to perform the desired intensity linear transformation $A$ between the input and the output FOVS, by directly using the relationship $o = A\,i$ to generate a large set of input-target intensity pairs $(i, o)$. Using this large training set of input/output intensity patterns, we optimized the transmission phase values of the diffractive layers using deep learning, by minimizing the MSE loss between the output pixel intensities of the diffractive processor $o'$ and the ground-truth intensities $o$ (see the Methods section). During the training phase, the output intensity of the diffractive processor was simulated through the incoherent propagation of the input intensity patterns, $i$ or $I(m',n')$. To numerically simulate the spatially-incoherent propagation of $I(m',n')$, we assumed the input optical field to be $\sqrt{I}e^{j\varphi}$ where $\varphi$ is a random 2D phase distribution, i.e., $\varphi(m',n') \sim Uniform(0, 2\pi)$ for each $(m',n')$. This input field with the random phase distribution $\varphi$ was coherently propagated through the diffractive surfaces to the output plane, using the angular spectrum approach[22]. We repeated this coherent wave propagation $N_\varphi$ times for every $i$, each time with a different random phase $\varphi(m',n')$ distribution, and averaged the resulting $N_\varphi$ output intensities. As $N_\varphi \to \infty$, the average intensity would approach the theoretical time-averaged output intensity for spatially-incoherent illumination, i.e., $O(m,n) = \langle |o(m,n)|^2 \rangle$. Due to the limited availability of computational resources, for the direct training (the second design approach) of the spatially-incoherent diffractive optical processors, we used $N_\varphi = N_{\varphi,tr} = 1000$.

The diffractive models reported in Figs. 1-5 and 10 were trained using the indirect approach while the ones in Figs. 6-9 were trained using the direct approach. All the diffractive networks reported in this work, after their training using either the direct or indirect design approaches, were evaluated and blindly tested through the incoherent propagation of input intensities with $N_{\varphi,te} = 20{,}000$. Since the testing is computationally less cumbersome compared to the training, we used $N_{\varphi,te} = 20{,}000 \gg N_{\varphi,tr}$.

After the training phase, we tested the resulting diffractive processor designs using 20,000 test intensity patterns $i$ that were never used during training; the size of this testing intensity set (20,000) should not be confused with $N_{\varphi,te} = 20{,}000$ since for each input intensity test pattern of this set, we used $N_{\varphi,te} = 20{,}000$ random 2D phase patterns to compute the corresponding spatially-incoherent output intensity. In Fig. 1c, the approximation errors of eight different phase-only diffractive processors trained using the indirect approach, each with $K = 5$ diffractive layers, are reported as a function of $N$. The mean error (Fig. 1c) for each diffractive design was calculated at the output intensity patterns $o'$ with respect to the ground truth $o = Ai$, by averaging over the 20,000 test intensity patterns. Figure 1c reveals that the approximation error of the spatially-incoherent diffractive processors reaches a minimum level as $\frac{N}{2N_iN_o}$ approaches 1, and stays at the same level for $N \geq 2N_iN_o$.

To understand the impact of $N_{\varphi,te}$ on these approximation error calculations, we took the diffractive processor design # 1E shown in Fig. 1c (i.e., $K = 5, N \approx 2.1 \times 2N_iN_o$), and used different $N_{\varphi,te}$ values at the blind testing phase for evaluating the average test error on the same intensity test set composed of 20,000 patterns $\boldsymbol{i}$. As shown in Fig. 1d, the computed error values decrease as $N_{\varphi,te}$ increases, as expected. On the right y-axis of the same Figure 1d, we also show, as a function of $N_{\varphi,te}$, the expectation value of $\left|\frac{1}{N_{\varphi,te}}\sum_{i=1}^{N_{\varphi,te}} e^{j\theta_i}\right|$, where $\theta_i \sim Uniform(0, 2\pi)$. This expectation value of the residual magnitude of $\frac{1}{N_{\varphi,te}}\sum_{i=1}^{N_{\varphi,te}} e^{j\theta_i}$ decreases as $N_{\varphi,te}$ increases and would approach zero as $N_{\varphi,te} \to \infty$. The numerically simulated output intensity of a diffractive processor design approaches the true time-averaged intensity of the spatially-incoherent wave as $N_{\varphi,te}$ gets larger, following a similar trend as $\left|\frac{1}{N_{\varphi,te}}\sum_{i=1}^{N_{\varphi,te}} e^{j\theta_i}\right|$, reported in Fig. 1d. This comparison also highlights the fact that our choice of using $N_{\varphi,te} = 20,000$ random 2D phase patterns to compute the spatially-incoherent output intensity patterns in the blind testing phase is an accurate approximation.

Next, we show in Fig. 2 the scaled intensity linear transformations, $\widehat{\boldsymbol{A}}$, that were approximated by five of the trained diffractive networks of Fig. 1c. $\widehat{\boldsymbol{A}}$ is related to the physical transformation $\boldsymbol{A}'$ by a scalar factor $\sigma_A$ (see the 'Evaluation' subsection in 'Methods' section) which compensates for diffraction efficiency-related optical losses. We also show the error matrix with respect to the target $\boldsymbol{A}$, i.e., $\boldsymbol{\varepsilon} = |\boldsymbol{A} - \widehat{\boldsymbol{A}}|$, and report the average of the error matrix elements in the table on the right. Here $|\cdot|$ denotes the elementwise operation. As $N$ increases, the diffractive networks' resulting matrices resemble the ground truth target better and the approximation error decreases steadily; however, the improvement is more prominent as $N$ approaches $2N_iN_o$ and stagnates beyond $N \approx 2N_iN_o$.

To provide visually more noticeable illustrations of the diffractive networks' all-optical intensity transformations under spatially-incoherent illumination, we used structured intensity patterns such as the letters U, C, L, and A as input intensity to the diffractive networks (see Fig. 3). Because of the randomness of the elements of the intensity transformation matrix, the output pixel intensities also appear random (harder to compare visually against the ground truth). However, the reappearance of the letters after a numerical inversion through the multiplication of the scaled output intensity $\widehat{\boldsymbol{o}}$ by the inverse of the target transformation, $\boldsymbol{A}^{-1}$, would indicate $\widehat{\boldsymbol{A}} \approx \boldsymbol{A}$ and validate the correctness of the diffractive networks' approximations in a visually noticeable manner (see the 'Evaluation' subsection of the Methods section for the definition of $\widehat{\boldsymbol{o}}$). In the case of the diffractive network # 1A ($K = 5, N = 5 \times 38^2 \approx 0.88 \times 2N_iN_o$), the result of such an inversion does not quite reveal any recognizable patterns, indicating the near-failure of the all-optical approximation of this design # 1A. However, such inversion reveals the recognizable patterns (U, C, L, and A) as $N$ approaches $2N_iN_o$ (design # 1B) and becomes identical to the inputs as $N$ exceeds $2N_iN_o$ (e.g., design # 1C). These results show that for the $K = 5$ phase-only diffractive networks with a sufficiently large $N \geq \sim 2N_iN_o$, we have $\widehat{\boldsymbol{A}} \approx \boldsymbol{A}$, indicating that these networks could faithfully approximate the target intensity linear transformation under spatially-incoherent illumination.

For computational imaging and sensing applications, such as in microscopy, exploring patterns of closely spaced lines and points would be interesting. Motivated by this, we repeated the same procedures outlined in Fig. 3 for various intensity patterns consisting of closely separated line pairs and sets of points, the results of which are summarized in Fig. 4. The same conclusions drawn previously in Fig. 3 hold: for $N \geq \sim 2N_i N_o$ we have $\widehat{A} \approx A$.

We also investigated the dependence of the all-optical approximation of intensity linear transformations on the number of diffractive layers $K$; see Fig. 5. The results of this analysis reveal that even with $N \approx 2 \times 2N_i N_o$, $K = 1$ and $K = 2$ diffractive designs failed to approximate the target linear transformation despite having a large $N$, whereas the designs with $K > 2$ successfully approximated the target transformation under spatially-incoherent illumination. This confirms that the depth of the diffractive network design is a key architectural factor in the computational capacity of diffractive processors to perform arbitrary linear transformations[21,22,37,38].

Next, we present the blind testing results of the diffractive processors that were trained using the second design approach (i.e., direct approach), to perform the same arbitrary intensity linear transformation as has been considered so far. In Fig. 6a, the approximation errors of eight different phase-only diffractive processors trained using the direct approach, each with $K = 5$ diffractive layers, are reported as a function of $N$. The mean error was calculated over the same 20,000 test intensity patterns used in Fig.1c; for each test intensity pattern, the incoherent output intensity $o'$ was calculated using $N_{\varphi,te} = 20000$ (same as before). In these alternative diffractive designs, the approximation error of the diffractive processors reaches a minimum level as $\frac{N}{2N_i N_o}$ approaches 1, and stays at the same level for $N \geq 2N_i N_o$ – the same conclusion that we reached for the indirect designs reported earlier. However, compared with the previous designs that used the indirect approach, here, the minimum error level obtained using the direct approach is approximately three times higher. This can be attributed to the use of a relatively small $N_{\varphi,tr} = 1000$ during the training, and these designs can be further improved by increasing $N_{\varphi,tr}$ using a longer training effort with more computational resources.

In Fig. 7, we show the scaled linear intensity transformations, $\widehat{A}$, that were approximated by five of the trained diffractive networks of Fig. 6a. For each case, we also show the error matrix with respect to the target $A$, i.e., $\varepsilon = |A - \widehat{A}|$, and report the average of the error matrix elements in the table on the right. As $N$ increases, the mean intensity transformation error decreases, except for design # 2B which we believe is an outlier resulting from poor convergence. The relatively large error of the design # 2B is due to the diffraction efficiency imbalance among the individual input pixels, as evident from the uneven magnitudes across the columns of $\widehat{A}$. Similarly, the other designs of the direct approach reveal uneven magnitudes across the columns of $\varepsilon$, indicating some diffraction efficiency imbalance among the individual input pixels, albeit not as severe as the design # 2B. Despite such imperfections, these diffractive networks designed using the direct approach effectively learned the target intensity transformation, as evident from Figs. 8 and 9. Figure 8 reveals that, for all the designs, the multiplication of the output intensity patterns $\widehat{o}$ by the inverse of the target transformation, $A^{-1}$ brings back the patterns U, C, L, A. Although, the reconstruction quality is better for $N \approx 2N_i N_o$ and remains similar beyond $N > 2N_i N_o$, the improvement is not as sharp as it was with the indirect approach (see Fig. 8 vs. Fig. 3 and Fig. 9 vs. Fig. 4). In contrast with the diffractive networks designed using the indirect approach, here in this case, the diffractive networks with $N < 2N_i N_o$ (e.g., design # 2A) succeeded in approximating the linear transformation to the extent of revealing recognizable patterns after a numerical inverse mapping. These same observations also hold for the intensity patterns that consist of closely spaced lines and points, as shown in Fig. 9.

Finally, we report in Fig. 10 the performance of a diffractive network ($K = 5$, $N \approx 2 \times 2N_i N_o$) trained using the indirect approach to approximate another arbitrary intensity linear transformation, defined by a non-invertible matrix. The target transformation $\boldsymbol{A}$, the approximate all-optical transformation $\widehat{\boldsymbol{A}}$, and the error matrix $\boldsymbol{\varepsilon} = |\boldsymbol{A} - \widehat{\boldsymbol{A}}|$ are shown in Fig. 10a, revealing that the diffractive network design performed the target intensity transformation with negligible error. We also show the performance of this diffractive network design on test patterns (U, C, L, and A as well as line pairs and points) in Fig. 10b. The all-optical outputs are identical to the ground truth outputs, further confirming that we have $\widehat{\boldsymbol{A}} \approx \boldsymbol{A}.$ Another example of the all-optical approximation of an arbitrary intensity transformation (defined by a random permutation matrix) is also reported in Supplementary Figure S1.

## Discussion

We demonstrated that phase-only diffractive networks under spatially-incoherent illumination could perform arbitrary linear transformations of optical intensity with a negligible error if $N \geq 2N_i N_o$. The same conclusions would be applicable to complex-valued diffractive networks where the phase and amplitude of each diffractive feature could be independently optimized; in that case, the critical number of complex-valued diffractive features for approximating an arbitrary linear transformation of optical intensity would reduce by half to $N_i N_o$ due to the increased degrees of freedom per diffractive layer. Because of the practical advantages of phase-only diffractive networks, without loss of generality, we limited our analyses in this work to phase-only modulation at each diffractive surface.

Our results suggest that the two different training approaches (indirect vs. direct design) converge differently. If $N$ is comparable to or larger than $2N_i N_o$, the indirect approach results in significantly better and faster convergence and accurate approximation $\widehat{\boldsymbol{A}} \approx \boldsymbol{A}$; on the other hand, the direct design approach works better when $N$ is considerably less than $2N_i N_o$, even if its approximation error is larger. For example, although the designs # 2A and # 2B have higher errors than the design # 1A, the performances of the former on various test patterns are manifestly better as compared in Figs. 3, 4, 8 and 9. These direct designs can be further improved in their approximation power by increasing $N_{\varphi,tr} \gg 1000$ through a longer training phase, utilizing more computational resources.

An important advantage of the direct approach over the indirect one is that the former is compatible with data-driven design and can be applied even if the only information available to the designer is the sample data representing the target incoherent linear process, without a priori knowledge of the transformation matrix itself. By the same token, the direct approach also lends itself to data-driven optimization of incoherent diffractive processors for all-optical linear approximation of some nonlinear processes. As a consequence of this, data-driven design of incoherent processors for performing other inference tasks such as e.g., all-optical image classification under spatially-incoherent illumination, can be accomplished using the direct approach.

The failure of shallow diffractive networks to perform an arbitrary intensity transformation (see e.g., $K = 1$ and $K = 2$ designs shown in Fig. 5) indicates that shallow architectures with phase-only diffractive layers are unable to effectively balance the ballistic photons that are transmitted from the sample/input FOV over a low numerical aperture; as a result of this, the lower spatial frequencies of the input intensity patterns dominate the output intensity patterns of a shallow diffractive network, sacrificing the approximation accuracy. Therefore, shallow diffractive network architectures, even with large numbers of trainable diffractive features ($N$), fail to approximate an arbitrary intensity transformation, as shown in Fig. 5. Deeper architectures, on the other hand, utilize their trainable diffractive features more effectively by distributing them across several layers/surfaces, one following another, and mixing the propagating modes of the input FOV over a series of layers that are optimized using deep learning.

Spatially-incoherent diffractive processor designs can also be extended to temporally incoherent broadband illumination light. In fact, multiplexing of >100 arbitrary complex-valued linear transformations for complex optical fields was shown to be possible under spatially-coherent but broadband illumination light[38]. Following a similar multi-wavelength optimization process and the indirect design principles outlined earlier, one can design a diffractive network to simultaneously approximate a group of arbitrarily-selected linear intensity transformations $(A_{\lambda_1}, A_{\lambda_2}, \ldots A_{\lambda_M})$ under spatially-incoherent illumination, where each intensity transformation is assigned to a unique wavelength $\lambda_i \{i = 1: M\}$. The success of such a spatially- and temporally-incoherent diffractive optical network to accurately perform all the target intensity transformations will require an increase in the number of trainable features within the diffractive volume, i.e., $N \geq M \times 2N_iN_o$ would be needed for a phase-only diffractive network. Such diffractive processor designs that work under spatially- and temporally-incoherent light can be useful for a number of applications, including fluorescence and brightfield microscopy and the processing of natural scenes.

## Methods
### Model for the propagation of spatially-coherent light through a diffractive optical network

Propagation of spatially-coherent complex optical fields through a diffractive processor $\mathfrak{D}\{\cdot\}$ constitutes successive amplitude and/or phase modulation by diffractive surfaces, each followed by coherent propagation through the free space separating consecutive diffractive surfaces. The diffractive features of a surface locally modulate the incident optical field $u(x, y)$. For this paper, the trainable diffractive features are *phase-only*, i.e., only the phase, but not the amplitude, of the incident field is modulated by the diffractive surface. In other words, the field immediately after the surfaces would be $u(x, y) \exp(j\phi_M(x, y))$ where the local phase change $\phi_M(x, y)$ induced by the surface is related to its height $h(x, y)$ as $\phi_M = \frac{2\pi}{\lambda}(n - 1)h$. Here $n$ is the refractive index of the diffractive surface material.

Free-space propagation of an optical field between consecutive diffractive surfaces was modeled using the angular spectrum method [8], according to which the propagation of an optical field $u(x, y)$ by distance $d$ can be computed as follows:

$$u(x, y; z = z_0 + d) = \mathcal{F}^{-1}\{\mathcal{F}\{u(x, y; z = z_0)\} \times H(f_x, f_y; d)\} \quad (6)$$

where $\mathcal{F}$ ($\mathcal{F}^{-1}$) is the two-dimensional Fourier (Inverse Fourier) transform and $H(f_x, f_y; d)$ is the free-space transfer function for an axial propagation distance $d$:

$$H(f_x, f_y; d) = \begin{cases} \exp\left(j\frac{2\pi}{\lambda}d\sqrt{1 - (\lambda f_x)^2 - (\lambda f_y)^2}\right), & f_x^2 + f_y^2 < 1/\lambda^2 \\ 0, & \text{otherwise} \end{cases} \quad (7)$$

where $\lambda$ is the wavelength of light.

## Model for the propagation of spatially-incoherent light through a diffractive optical network

With spatially-incoherent light, the (average) output optical intensity $O(x,y)$ of a diffractive network, for a given input intensity $I(x,y)$, can be written as

$$O(x,y) = \langle |\mathfrak{D}\{\sqrt{I(x,y)}\exp(j\varphi(x,y))\}|^2 \rangle = \lim_{N_\varphi \to \infty} \frac{1}{N_\varphi} \sum_{r=1}^{N_\varphi} |\mathfrak{D}\{\sqrt{I(x,y)}\exp(j\varphi_r(x,y))\}|^2 \quad (8)$$

where $\mathfrak{D}\{\cdot\}$ denotes the coherent propagation of the optical field through the diffractive processor as described in the preceding subsection, and $\langle \cdot \rangle$ denotes the statistical average, over all the realizations of the spatially-independent random process $\varphi(m,n)$ representing the 2D phase of the input optical field, i.e., $\varphi(m,n) \sim U(0, 2\pi)$ for all $m,n$[40].

As for the spatially-incoherent propagation of average intensity, it is only possible to approximate the true average (Eq. 8) by averaging over a finite number $N_\varphi$ of samples of $\varphi(x,y)$, i.e.,

$$O(x,y) \approx \frac{1}{N_\varphi} \sum_{r=1}^{N_\varphi} |\mathfrak{D}\{\sqrt{I(x,y)}\exp(j\varphi_r(x,y))\}|^2 \quad (9)$$

In the training phase of the direct training approach, incoherent propagation of intensities through the diffractive processors was simulated with $N_{\varphi,tr} = 1000$. However, in the blind testing phase we used $N_{\varphi,te} = 20000$ while evaluating the diffractive processors once they were trained, irrespective of whether the indirect or the direct approach of training was used.

In our numerical simulations, the fields/intensities were discretized using $\delta \approx 0.53\lambda$ along both $x$ and $y$, e.g., $u(m,n) \triangleq u(m\delta, n\delta)$ and sufficiently zero-padded before evaluating the Fourier transform, as in Eq. 6, using Fast Fourier Transform (FFT) algorithm.

## Diffractive network architecture

The heights $h(m,n) \triangleq h(m\delta, n\delta)$ of the $N$ diffractive features distributed over $K$ surfaces were optimized for designing the diffractive processors to perform the desired transformation. To keep the connectivity between successive diffractive layers[22] the same across the trained diffractive networks with different $N$, the layer-to-layer separation was set as $d = \frac{W\delta}{\lambda}$, where $W = \sqrt{\frac{N}{K}}\delta$ is the width of each diffractive layer. The distances between the input FOV and layer-1 and between layer-$K$ and the output FOV were also set as $d$. The pixel size on both the input and the output FOVs was ~$2.13\lambda \times 2.13\lambda$, i.e., $4\delta \times 4\delta$.

## Linear transformation matrix

In this paper, the input and the output of the diffractive networks have dimensions of $N_i = N_o = 8 \times 8$, i.e., $I, O \in \mathbb{R}_+^{8 \times 8}$ and $\boldsymbol{i}, \boldsymbol{o} \in \mathbb{R}_+^{64}$. To clarify, $\boldsymbol{i}$ and $\boldsymbol{o}$ are one-dimensional (column) vectors obtained by rearranging the intensity values $I(m,n)$ and $O(m,n)$ of the input and the output pixels arranged in a two-dimensional $8 \times 8$ square grid. Accordingly, the target transformation matrix $\boldsymbol{A}$ has a size of $N_o \times N_i = 64 \times 64$, i.e., $\boldsymbol{A} \in \mathbb{R}_+^{64 \times 64}$.

### Training details

The height $h$ of the diffractive features at each layer was confined between zero and a maximum value $h_{max}$ by using a latent variable $h_{latent}$:

$$h = \frac{h_{max}}{2} \times [\sin(h_{latent}) + 1]$$

We chose $h_{max} \approx \frac{\lambda}{n-1}$ so that the corresponding phase modulation depth is $2\pi$. The diffractive layers were optimized using the AdamW optimizer [41] for 50 epochs with a minibatch size of 8 and an initial learning rate of $10^{-3}$. The learning rate was decayed by a factor of 0.7 every five epochs. The latent variables were initialized randomly from the standard normal distribution $\mathcal{N}(0,1)$. We evaluated the mean loss of the trained model on the validation set after the completion of each epoch and selected the trained model state at the end of the epoch corresponding to the lowest validation loss. These details were the same for both the indirect and the direct training approaches.

The diffractive processor models were implemented and trained using PyTorch (v1.10) [42] with Compute Unified Device Architecture (CUDA) version 11.3.1. Training and testing were done on GeForce RTX 3090 graphics processing units (GPU) in workstations with 256GB of random-access memory (RAM) and Intel Core i9 central processing unit (CPU). The training time of the models varied with the training approach as well as the size of the models in terms of $K$ and $N$. For example, the indirect training of $K = 5, N = 5 \times 52^2$ diffractive network model took around 5 hours , whereas with the direct approach, the training time for the $K = 5, N = 5 \times 52^2$ model with $N_{\varphi,tr} = 1000$ was around 58 hours.

### Evaluation

The evaluation procedure was the same across all the trained diffractive networks irrespective of whether the direct approach or the indirect approach was used to train them. To evaluate the trained diffractive networks, we generated a test set comprising 20,000 pairs of input and target intensity vectors $\boldsymbol{o} = \boldsymbol{Ai}$. Note that these 20,000 test examples were generated using a different random seed from the ones used to generate the training and the validation sets to ensure they were not represented during the training. For a given $\boldsymbol{i}$, the corresponding input intensity pattern was incoherently propagated through the trained diffractive network (as in Eq. 9) using $N_{\varphi,te} = 20,000$ to compute the output intensity $\boldsymbol{o}'$. The mean of the error between $\boldsymbol{o}'$ and $\boldsymbol{o}$ over the 20,000 test examples was used to quantify the output error of the diffractive network for comparing different designs, as in Figs. 1 and 6. For comparison between the ground truth and the all-optical output intensities, e.g., in Figs. 3, 4, 8, 9, 10, we defined the scaled all-optical output intensity vector $\hat{\boldsymbol{o}} = \frac{\sigma'}{\sigma}\boldsymbol{o}'$ (see Supplementary Information for details).

For evaluating the intensity transformation $\boldsymbol{A}'$ performed by the diffractive networks at the end of their training, we used $N_i$ intensity vectors $\{\boldsymbol{i}_t\}_{t=1}^{N_i}$ where $\boldsymbol{i}_t[l] = 1$ if $l = t$ and 0 otherwise. In other words, $\{\boldsymbol{i}_t\}_{t=1}^{N_i}$ are unit impulse functions where the impulses are located at different input pixels. We simulated the all-optical output intensity vectors $\{\boldsymbol{o}'_t\}_{t=1}^{N_i}$ corresponding to these input intensity vectors by incoherent propagation and stacked them column by column, i.e.,

$$\boldsymbol{A}' = [\boldsymbol{o}'_1|\boldsymbol{o}'_2|\cdots|\boldsymbol{o}'_{N_i}] \tag{10}$$

Considering the diffraction-efficiency-associated scaling mismatch between $A'$ and the target transformation $A$, we defined the scaled diffractive network intensity transformation $\widehat{A} = \sigma_A A'$, where:

$$\sigma_A = \sqrt{\frac{\sum_{q=1}^{N_i} \sum_{p=1}^{N_o} (A[p,q])^2}{\sum_{q=1}^{N_i} \sum_{p=1}^{N_o} (A'[p,q])^2}} \tag{11}$$

## Supplementary Information includes:
- The indirect approach of training
- The direct approach of training
- Supplementary Figure S1

**Figures and Figure captions:**

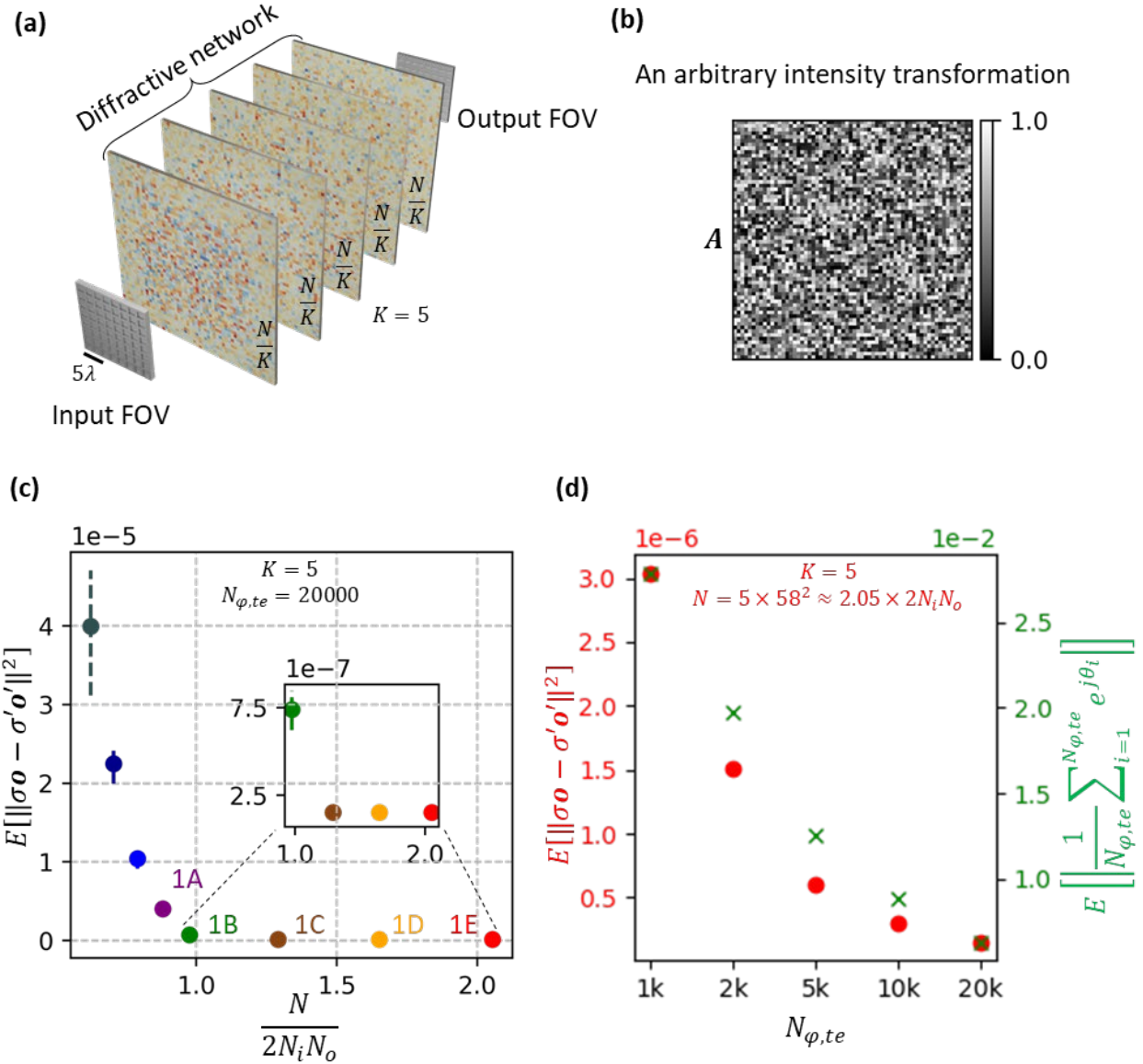

**Fig. 1:** All-optical linear transformation of intensity performed by diffractive networks under spatially-incoherent illumination. (a) Schematic of a diffractive network formed by $K = 5$ diffractive surfaces that all-optically perform a linear transformation of intensity between the input and output FOVs. The $N$ diffractive features are distributed evenly among the $K = 5$ surfaces. (b) An arbitrary $N_o \times N_i$ matrix $A$, representing the target intensity transformation to be performed all-optically by the diffractive network. Here $N_i = 8^2$ and $N_o = 8^2$ are the number of pixels at the input and the output FOVs of the diffractive network, respectively. (c) The expectation value of the MSE between the all-optical output intensity $o'$ and the ground-truth output intensity $o$, as a function of $N$ for different diffractive networks trained using the *indirect* approach. To simulate the incoherent propagation of intensity for each test input, we used $N_{\varphi,te} = 20000$. (d) Dependence of the calculated output MSE on $N_{\varphi,te}$, demonstrated for network

# 1E of Fig. 1c. The right y-axis shows the expectation value of the residual magnitude of $\frac{1}{N_{\varphi,te}}\sum_{i=1}^{N_{\varphi,te}} e^{j\theta_i}$, where $\theta_i \sim Uniform(0, 2\pi)$.

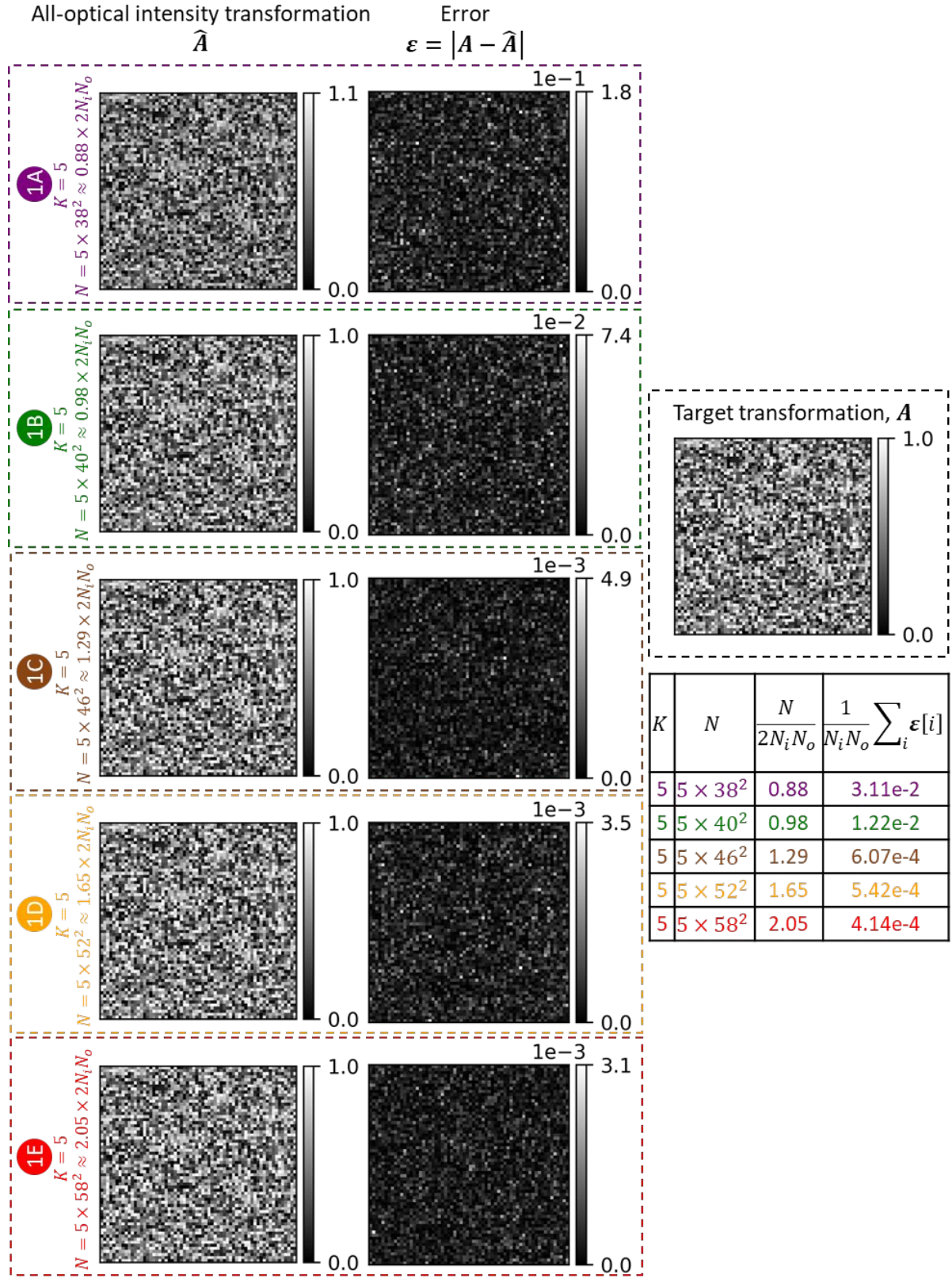

**Fig. 2:** All-optical linear transformations of intensity, $\widehat{A}$, performed under spatially-incoherent

illumination, by five of the diffractive network designs shown in Fig. 1c, together with the corresponding error matrices with respect to the target transformation, $\boldsymbol{\varepsilon} = |\boldsymbol{A} - \widehat{\boldsymbol{A}}|$. Here $|\cdot|$ denotes elementwise operation. The means of the error matrix elements are listed in the table on the right.

**Fig. 3:** All-optical linear transformation of structured intensity patterns such as letters U, C, L, and A by the same diffractive networks as in Fig. 2, accompanied by the patterns resulting from the numerical

inverse mapping of the all-optical outputs through multiplication by $A^{-1}$.

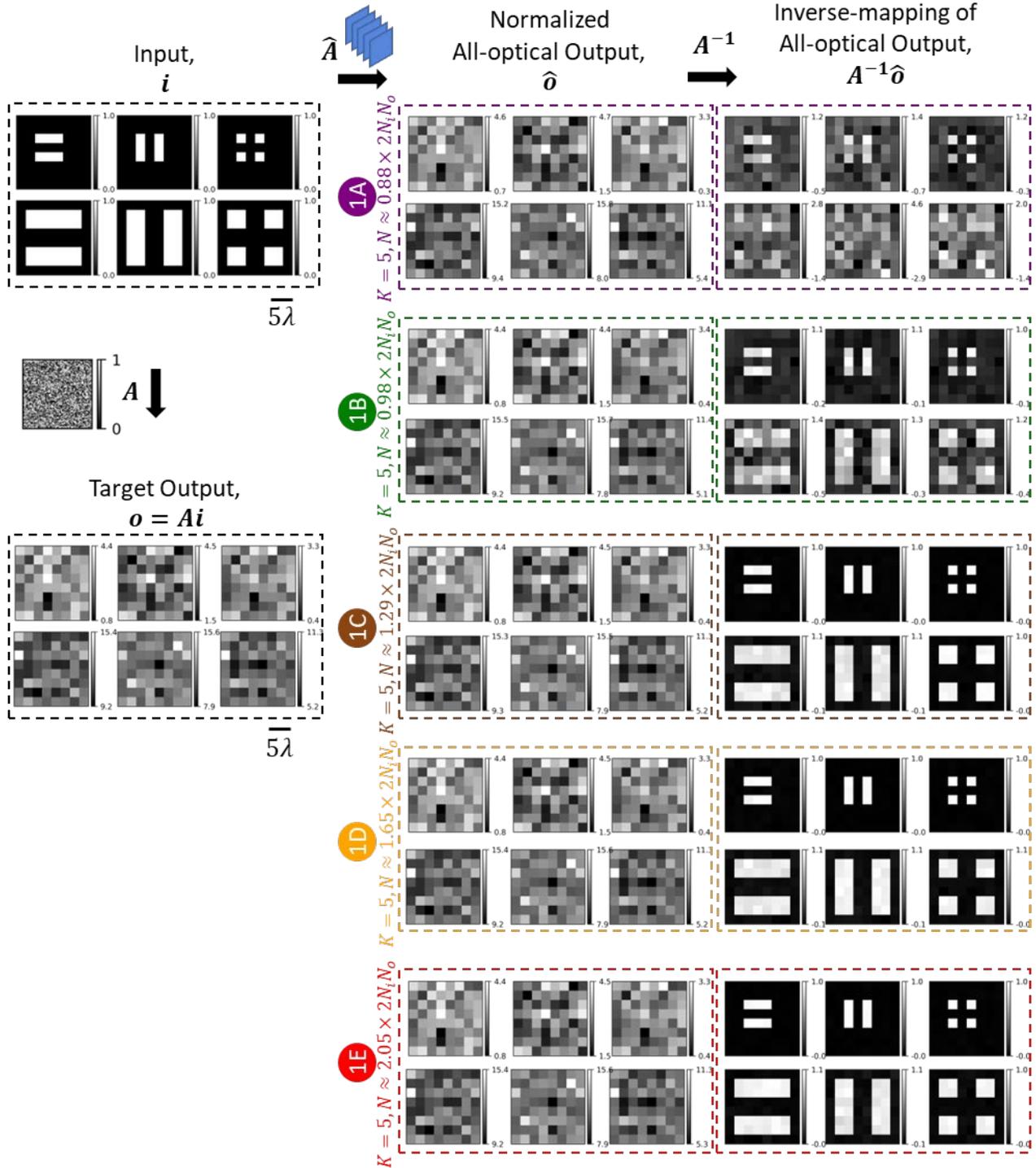

**Fig. 4:** Same as Figure 3, except for the test intensity patterns formed by closely separated lines and points.

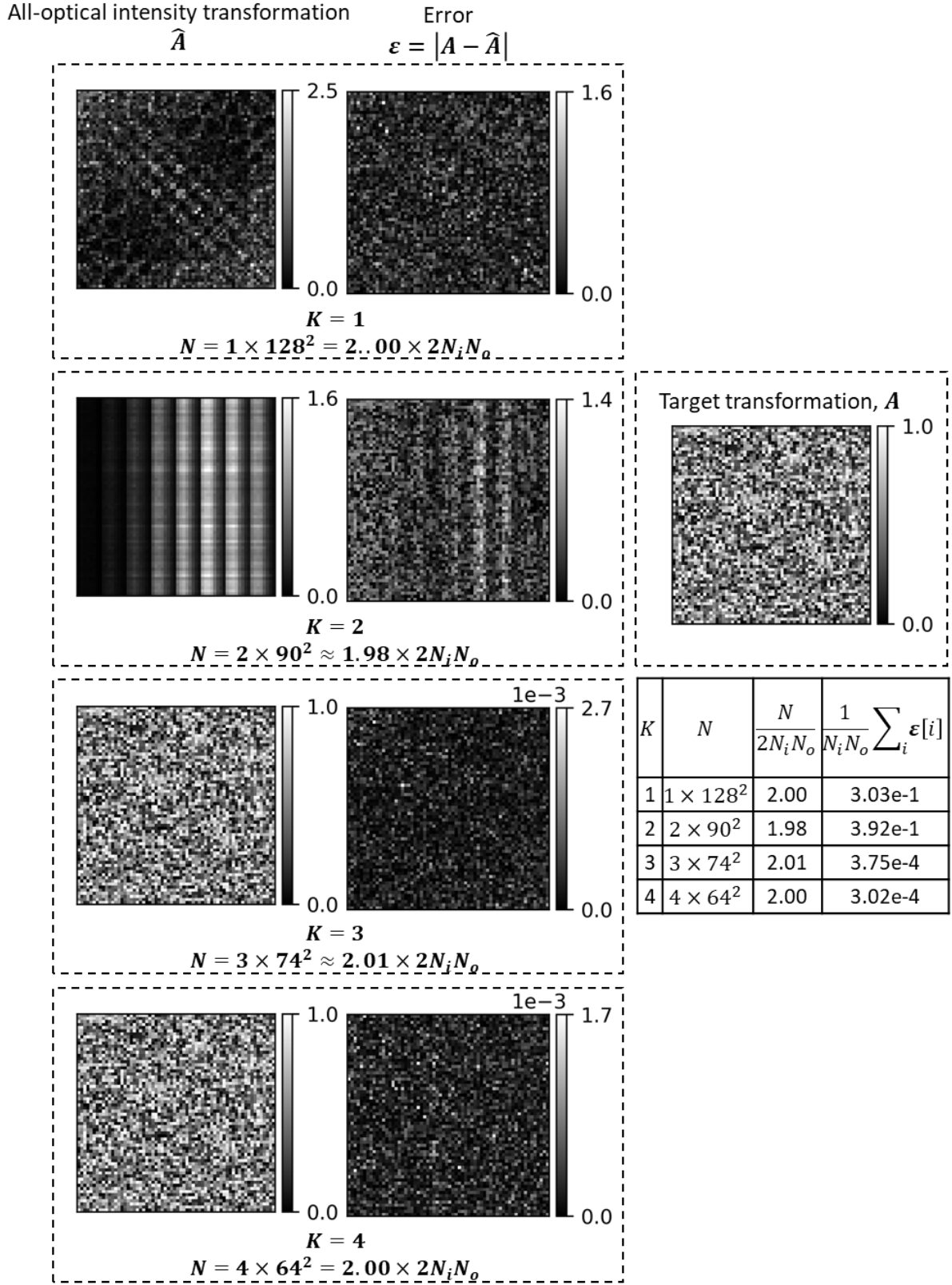

**Fig. 5:** Effect of the diffractive network's depth, i.e., the number of diffractive surfaces ($K$), on the approximation performance for an arbitrary intensity linear transformation under spatially-incoherent illumination. All-optical linear transformations of intensity, $\hat{A}$, performed by four diffractive network designs with approximately equal $N$ and increasing $K$, are shown, together with the corresponding error matrices with respect to the target transformation, i.e., $\varepsilon = |A - \hat{A}|$. Here $|\cdot|$ denotes elementwise operation. The mean values of the error matrix elements are listed in the table on the right.

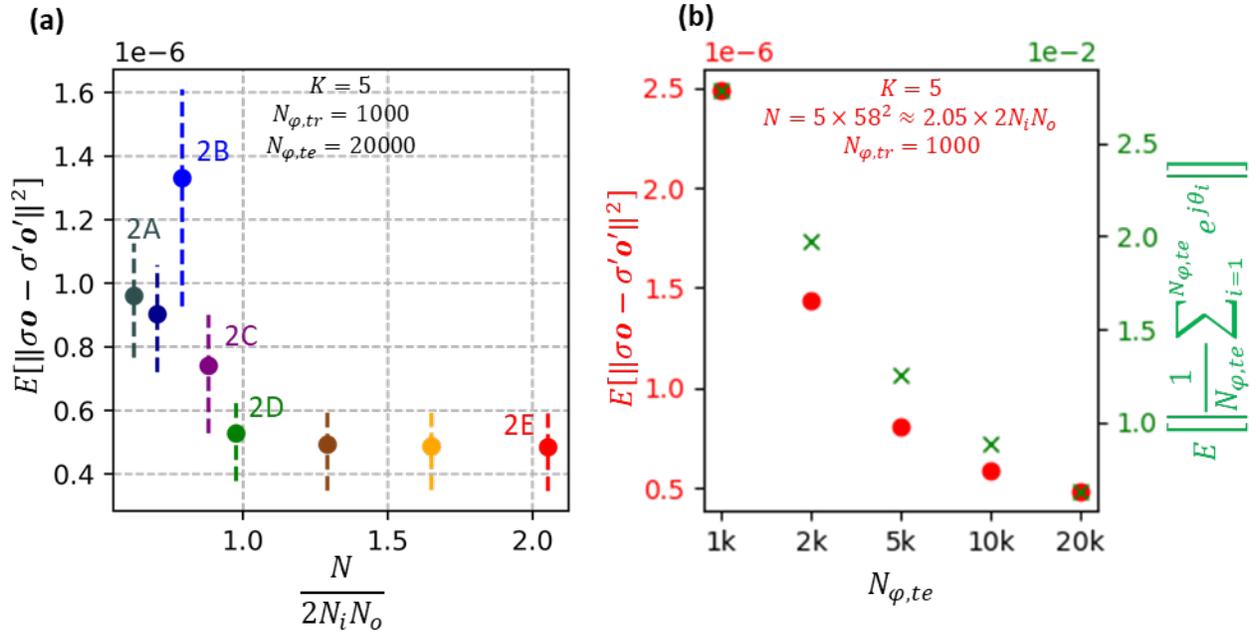

**Fig. 6:** All-optical linear transformation of intensity under spatially-incoherent illumination, by diffractive networks trained using the *direct* approach. (a) The expectation value of the MSE between the all-optical output intensity $o'$ and the ground-truth output intensity $o$, as a function of $N$ for different diffractive networks trained using the direct approach. (b) Dependence of the calculated output MSE on $N_{\varphi,te}$, demonstrated for network # 2E of Fig. 6a. The right y-axis shows the expectation value of the residual magnitude of $\frac{1}{N_{\varphi,te}} \sum_{i=1}^{N_{\varphi,te}} e^{j\theta_i}$, where $\theta_i \sim Uniform(0, 2\pi)$.

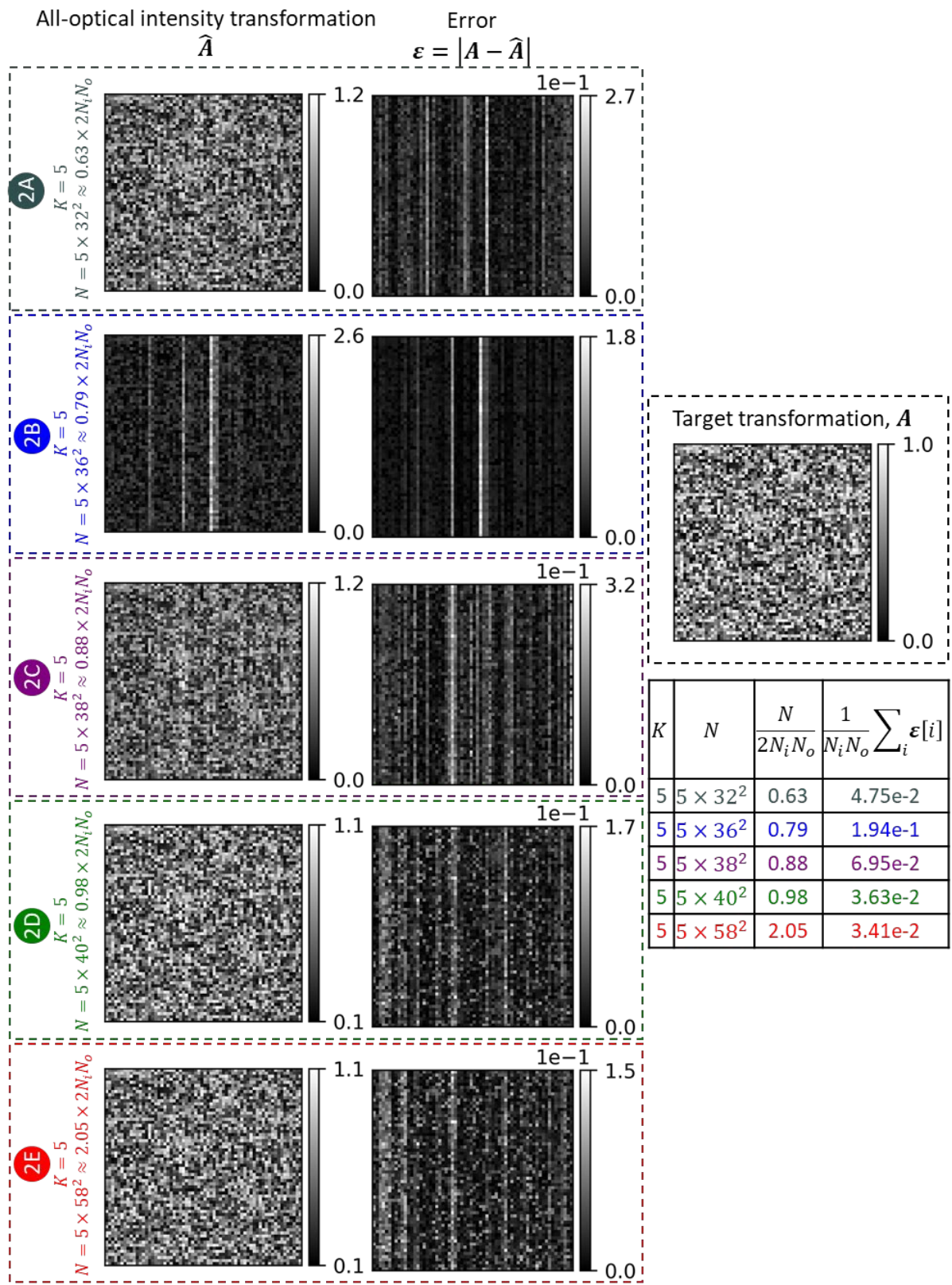

**Fig. 7:** All-optical linear transformations of intensity, $\widehat{A}$, performed under spatially-incoherent

illumination by five of the diffractive network designs shown in Fig. 6a, together with the corresponding error matrices with respect to the target transformation, $\varepsilon = |A - \widehat{A}|$. Here $|\cdot|$ denotes elementwise operation. The mean values of the error matrix elements are listed in the table on the right.

**Fig. 8:** All-optical linear transformation of structured intensity patterns such as letters U, C, L, and A by the same diffractive networks as in Fig. 7, accompanied by the patterns resulting from the numerical

inverse mapping of the all-optical outputs through multiplication by $A^{-1}$.

**Fig. 9:** Same as Figure 8, except for the test intensity patterns formed by closely separated lines and points.

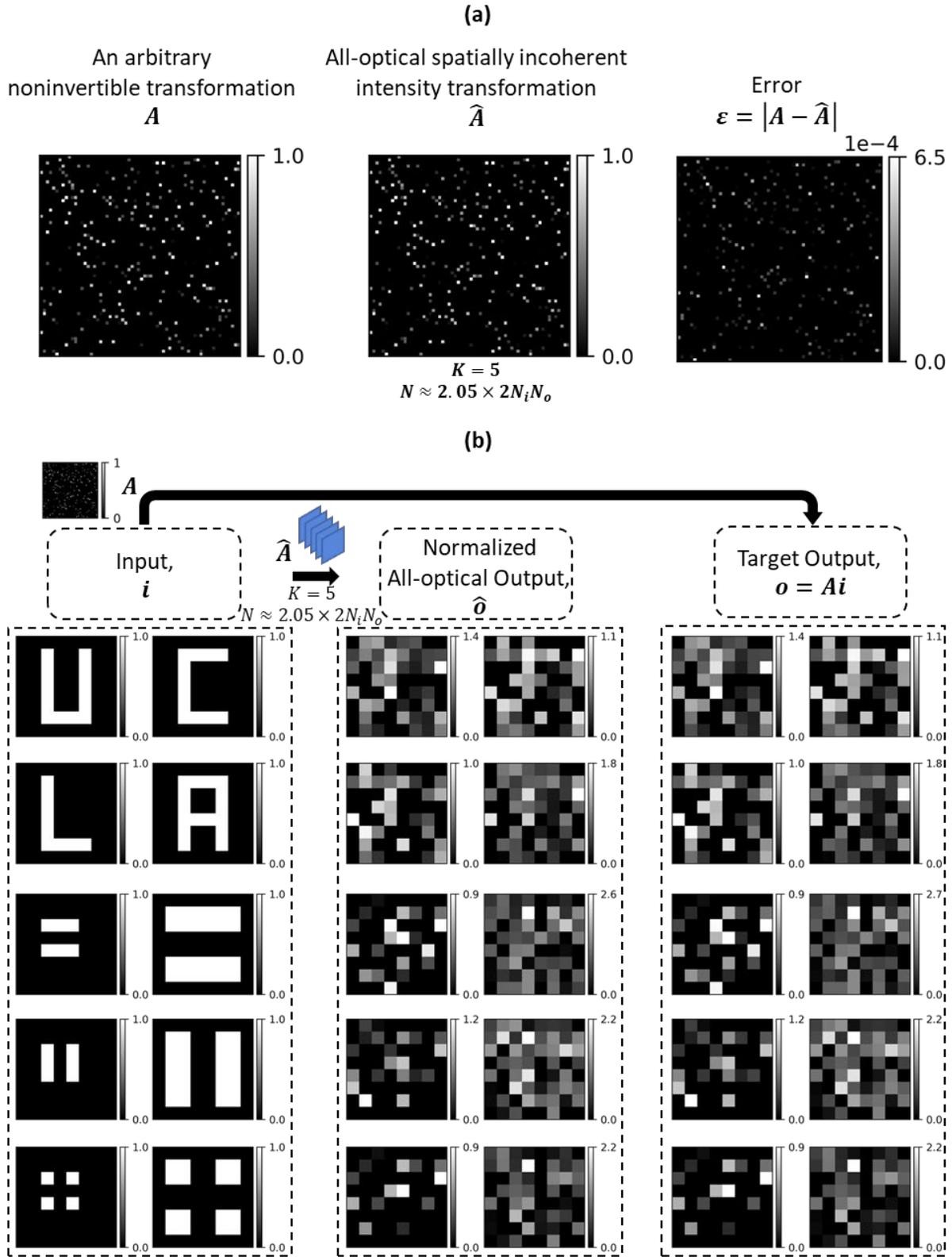

**Fig. 10:** Approximation of an arbitrary non-invertible linear transformation ($A$) of intensity, under spatially-incoherent illumination, by a diffractive network ($K = 5$, $N = 5 \times 58^2$) trained using the indirect approach. (a) The target transformation $A$, the all-optical intensity transformation $\widehat{A}$ performed by the trained diffractive network and the error matrix $\varepsilon = |A - \widehat{A}|$. Here $|\cdot|$ denotes elementwise operation. (b) All-optical transformation of different test intensity patterns by the trained diffractive network, together with the corresponding ground truths.